\documentclass[intlimits,twoside,a4paper]{article}

\usepackage{wrapfig}
\usepackage{xcolor}
\usepackage{graphicx}
\usepackage[T2A]{fontenc}
\usepackage[cp1251]{inputenc}

\usepackage{slantsc}

\usepackage{amsmath}
\usepackage{amsfonts}
\usepackage{amssymb}
\usepackage{epsfig}
\usepackage{xcolor}
\usepackage[normalem]{ulem}

\usepackage{cmpj2}
\issue{2016}{19}{1}{13804}
\doinumber{10.5488/CMP.19.13804}

\title[Anisotropic ion shape]
{Influence of anisotropic ion shape, asymmetric valency, and electrolyte concentration on structural and thermodynamic properties of an electric \\ double layer}

\author[M. Kaja \textsl{et al}.]{M. Kaja\refaddr{label1},
S. Lamperski\refaddr{label1}\thanks{E-mail: slamper@amu.edu.pl}\,,
W. Silvestre-Alcantara\refaddr{label3},
L.B. Bhuiyan\refaddr{label3},
D. Henderson\refaddr{label5}}

\addresses{
\addr{label1} Department of Physical Chemistry, Adam Mickiewicz University of Pozna\'n, \\ Umultowska 89b, 61-614 Pozna\'n, Poland
\addr{label3} Laboratory of Theoretical Physics, Department of Physics, University of Puerto Rico, \\ San Juan, 00931-3343, Puerto Rico
\addr{label5} Department of Chemistry and Biochemistry, Brigham Young University, Provo UT 84602-5700, USA
}

\date{Received November 20, 2015, in final form January 12, 2016}
\authorcopyright{M. Kaja, S. Lamperski, W. Silvestre-Alcantara,
L.B. Bhuiyan, D. Henderson, 2016}

\begin{document}
\maketitle

\begin{abstract}

Grand canonical Monte Carlo simulation results are reported for an electric double layer modelled by a planar charged hard wall, anisotropic shape cations, and spherical anions at different electrolyte concentrations and asymmetric valencies.
The cations consist of two tangentially tethered hard spheres of the same diameter, ${d}$.
One sphere is charged while the other is neutral.
Spherical anions are charged hard spheres of diameter ${d}$.
The ion valency asymmetry 1:2 and 2:1 is considered, with
the ions being immersed in a solvent mimicked by a continuum dielectric medium at standard temperature.
The  simulations are carried out for the following electrolyte concentrations: 0.1, 1.0 and 2.0~M.
Profiles of the electrode-ion, electrode-neutral sphere singlet distributions, the average orientation of dimers, and the mean electrostatic potential are calculated for a given electrode surface charge, $\sigma $, while the contact
electrode potential and the differential capacitance are presented for varying electrode charge.
With an increasing electrolyte concentration, the shape of differential capacitance curve changes from that with a minimum surrounded by maxima into that of a distorted single maximum.
For a 2:1 electrolyte, the maximum is located at a small negative $\sigma $ value while for 1:2, at a small positive  value.

\pacs 82.45.Fk, 61.20.Qg, 73.30.+y, 82.45.Mp, 61.20.Ja

\end{abstract}

This article is dedicated to Professor Stefan Soko{\l}owski, the famous Polish scientist and our friend, on the occasion of his 65th birthday.

\section{Introduction}

Development of the electrical double layer (EDL) theory as well as the progress in the numerical technology have resulted in proposition and examination of an increasing number of EDL models.
The Gouy-Chapman theory (GC) \cite{1.,2.} is based on solving the Poisson-Boltzmann equation and applies a mean field approximation to describe the electrostatic interactions.
The GC theory can be applied to the simplest model of an electrolyte, which  does not take into account the volume of ions.
In this model, ions are represented by point electric charges, and the solvent is a dielectric continuum with the relative permittivity $\epsilon_\text{r}$.
It is worth noting that the Debye-H\"{u}ckel theory uses the same model of an electrolyte.
The GC theory describes well low density electrolytes at small electrode charges.
It breaks down at higher concentrations and charges because the excluded volume and correlation effects are disregarded.

The hard-sphere and electrostatic correlation effects are considered in modern theories of EDL such as the Modified Poisson-Boltzmann Theory (MPB) \cite{3.,4.}, the Mean Spherical Approximation (MSA) \cite{5.}, the Hypernetted Chain Theory (HNC) \cite{6.,7.} or more recently the Density Functional Theory (DFT) \cite{8.}.
They have been used to describe the primitive model of an electrolyte (PM) \cite{9.} and the restricted primitive model (RPM) \cite{10.}.
In the PM model, ions are represented by hard spheres of different diameters with a point electric charge located at the centre.
The charged spheres are immersed in a medium with the relative permittivity $\epsilon_\text{r}$ characteristic of a solvent.
In the RPM model, the ion diameters are the same.
For EDL composed of a planar electrode and a PM or RPM electrolyte, the modern theories predict the layering effect at high absolute values of the electrode charges \cite{11.,12.} and the transition of the capacitance minimum into a maximum at small electrode charges caused by an increasing electrolyte concentration \cite{4.}.
The GC theory fails to describe these effects.

Due to correlation effects that are included, the MPB and DFT theories have been successfully applied to the solvent primitive model (SPM) electrolyte.
SPM is the simplest model of an electrolyte which includes the volume of solvent molecules.
In SPM, solvent molecules are modelled by neutral hard spheres whose diameter is the same  as \cite{13.,14.} or different  \cite{15.}  from that of ions.
Neutral spheres as well as cations and anions are immersed in the continuous dielectric medium characterised by $\epsilon_\text{r}$.
The presence of neutral spheres leads to the generation of density oscillations near the electrode surface and to an increase in the differential capacitance of EDL \cite{16.}.

Of the formal statistical mechanical theories, the HNC has been applied to the non-primitive model (NP) \cite{17.} in which solvent molecules are modelled by hard spheres with a point permanent electric dipole moment located at the centre. However, it leads to lowered values of the relative permittivity.

The DFT theory gives fresh possibilities.
Due to the free energy term of intra-molecular interactions, the DFT theory is applicable to ions of more complicated topology than spherical.
Here, we must mention the work by Soko{\l}owski \cite{18.} who has applied the modified DFT theory to the study of orientation ordering of electrostatic neutral hard dumbbells at the planar surface.
Charged dumbbells, called dimers, have aroused great interest in investigation of the EDL properties \cite{19.,20.}.
This advanced model of an electrolyte has been used for modelling systems of high densities like ionic liquids \cite{21.}.

Torrie and Valleau \cite{10.} have introduced the computer simulation technique into the investigation of EDL.
At that time computer simulation results were used to confirm the correctness of EDL theories.
Now, the tremendous development of numerical technology has opened new area of application inaccessible for theory.
Fedorov et al. \cite{22.} have studied models of ionic liquids made  of one, two or three beads, assuming that one of the hard spheres in the chain had a positive charge.
Breitsprecher et al. \cite{23.,24.} have conducted molecular dynamics simulations for ions with different size and valency, represented by a coarse-grained model.
Silvestre-Alcantara et al. \cite{25.} have investigated the properties of ELD containing fused dimer electrolyte.
Charged hard walls and hard spheres have been replaced by molecular electrodes and soft sphere ions \cite{26.,27.,28.}.
Ions of topology characteristic of ionic liquids are investigated.
The solvent is no longer a dielectric continuum but is modelled by explicit molecular models \cite{29.}.
Thus, the present models of EDL have become more realistic.

Recently, we have intensively investigated the EDL containing a dimer electrolyte.
Among other effects, we have analysed the influence of concentration of 1:1 electrolyte \cite{21.}  and of ion valence asymmetries \cite{30.}  of charged dimers on the properties of EDL.
However, we did not consider the properties of EDL for asymmetric ion valencies at different electrolyte concentrations.
In particular, we expect new shapes of the differential capacitance.
{Thus, in this paper, we discuss the influence of anisotropic ion shape on the structural and thermodynamic properties of ELD  containing asymmeteric 2:1 and 1:2 dimer electrolytes of different concentrations.}

\section{Model and methods}

The electric double layer is composed of a planar electrode and an  electrolyte, which is a mixture of spherical anions and anisotropic cations in the shape of dimers.
The dimer consists of two tangentially tethered hard spheres, one of which has a point electric charge immersed at the centre and the other is neutral.
The diameters of the two spheres of a dimer and the sphere of a monomer anion are the same.
The ion valencies are asymmetric with
the ions being immersed in a homogeneous medium of the relative permittivity, $\epsilon _\text{r}$.
The electrode is modelled by a hard planar wall with a uniformly distributed charge of surface density, $\sigma $.
The image effect is not considered which means that $\epsilon _\text{r}$ of the electrode material and solvent are the same.
The ion-ion and electrode-ion interactions are given by
\begin{equation}
 u_{ij}(r)=\left\{
\begin{array}{ll}
 \infty, & \quad r<(d_{i}+d_{j})/2, \\
\frac{1}{4\pi \epsilon _{0}\epsilon _\text{r}}\frac{e^{2} Z_i Z_j}{r}, & \quad r \geqslant (d_{i}+d_{j})/2,
\end{array}
\right.
\end{equation}
and
\begin{equation}
 u_{wi}(x)=\left\{
\begin{array}{ll}
 \infty,  & \quad x<d/2, \\
-\frac{\sigma Z_s ex}{\epsilon _{0}\epsilon _\text{r}}, & \quad x \geqslant d/2,
\end{array}
\right.
\end{equation}
respectively.
Here, $e$ is the magnitude of the elementary charge, and $Z_{s}$ is the valency of the particle of species $s$.
Also, $\epsilon _{0}$ is the vacuum permittivity,  $r$ is the separation between the centres of the two hard spheres, and $x$ is the perpendicular distance from the electrode surface to the centre of a hard sphere.

{The local number density $\rho_s(x)$ of the species $s$ at a distance $x$ is the first average quantity obtained from our simulations.
The reduced local density or the singlet distribution function $g_{s}(x)=\rho_s(x)/\rho_s^{0}$, ($\rho_s^{0}$ is the corresponding bulk number density) is used to describe the structure of ELD.
Also, the mean electrostatic potential $\psi (x)$ is defined in terms of $\rho_s(x)$}
\begin{equation}
\psi(x)=\frac{e}{\epsilon_0\epsilon _\text{r}}\sum_{s}Z_s\int_x^{\infty}\rho_s(x{'})(x-x{'})\rd x{'}.
\label{eq3}\end{equation}

The differential capacitance $C_\text{d}$ is the property which can be compared with the experimental results.
The differential capacity is defined as
\begin{equation}
C_\text{d} = \rd\sigma /\rd\psi (0).
\end{equation}
In practice, $C_\text{d}$ was calculated from the interpolation polynomials method introduced by Lamperski and Zydor \cite{31.}.

{The local (volume) charge density $\rho_{Q} (x)$
\begin{equation}
\rho_{Q}(x)=\sum\limits_{s} Z_{s}e g_{s}(x)
\end{equation}
and the local net charge per unit area, $\sigma_{\Sigma} (x)$
\begin{equation}
\sigma_{\Sigma}(x)=\sigma + \int_0^{x}\rho_Q(x{'})\rd x{'}
\end{equation}
are the properties that can be used for indicating the charge inversion (CI) and charge reversal (CR) phenomena.
The CI effect takes place when the electrode charge and the charge density of the second layer and, less commonly, subsequent layers of ions next to the electrode are of the same sign.
When the charge of the first layer of counter-ions overcomes the charge of an electrode, the electric field reverses its direction.
The function $\sigma_{\Sigma}(x)$ has the sign opposite to that of $\sigma$.
This effect is called the charge reversal \cite{32., 33.}.
Essentially at some $x$, the integrated charge overcompensates or overscreens the electrode charge.
Hence, the often used term \emph{charge overscreening} occurs in the literature.}

The second average quantity obtained from our simulations is the mean orientation function  $\langle \cos \theta \rangle$.
It depends on the distance ${x}$ from the electrode surface.
The function $\langle \cos \theta \rangle$ is the average value of $\cos \theta$, where $\theta$ is the angle between the normal to the electrode surface and the straight line joining the centres of hard spheres constituting a dimer.
The origin is located at the centre of the charged sphere.

The GCMC technique was applied to calculate the local densities, $\rho_s(x)$, and orientation $\langle \cos \theta \rangle$ profiles.
This technique is recommended for inhomogeneous systems \cite{34.} as it eliminates difficulties associated with the determination of the bulk concentration which appears when MC simulations are carried out in the canonical ensemble.
The details of the GCMC technique and its implementation to our investigation have been described in the previous papers \cite{4.,35.}.
The procedure FLIP3 \cite{36.} was used to rotate the dimer while the long-range electrostatic interactions were estimated with the method proposed by Torrie and Valleau  \cite{10.}.
The ionic activity coefficients required by the GCMC technique as input were obtained from the inverse GCMC method \cite{37.}.

\section{Results}

\begin{figure}[!b]
\centerline{
\includegraphics[width=0.5\textwidth]{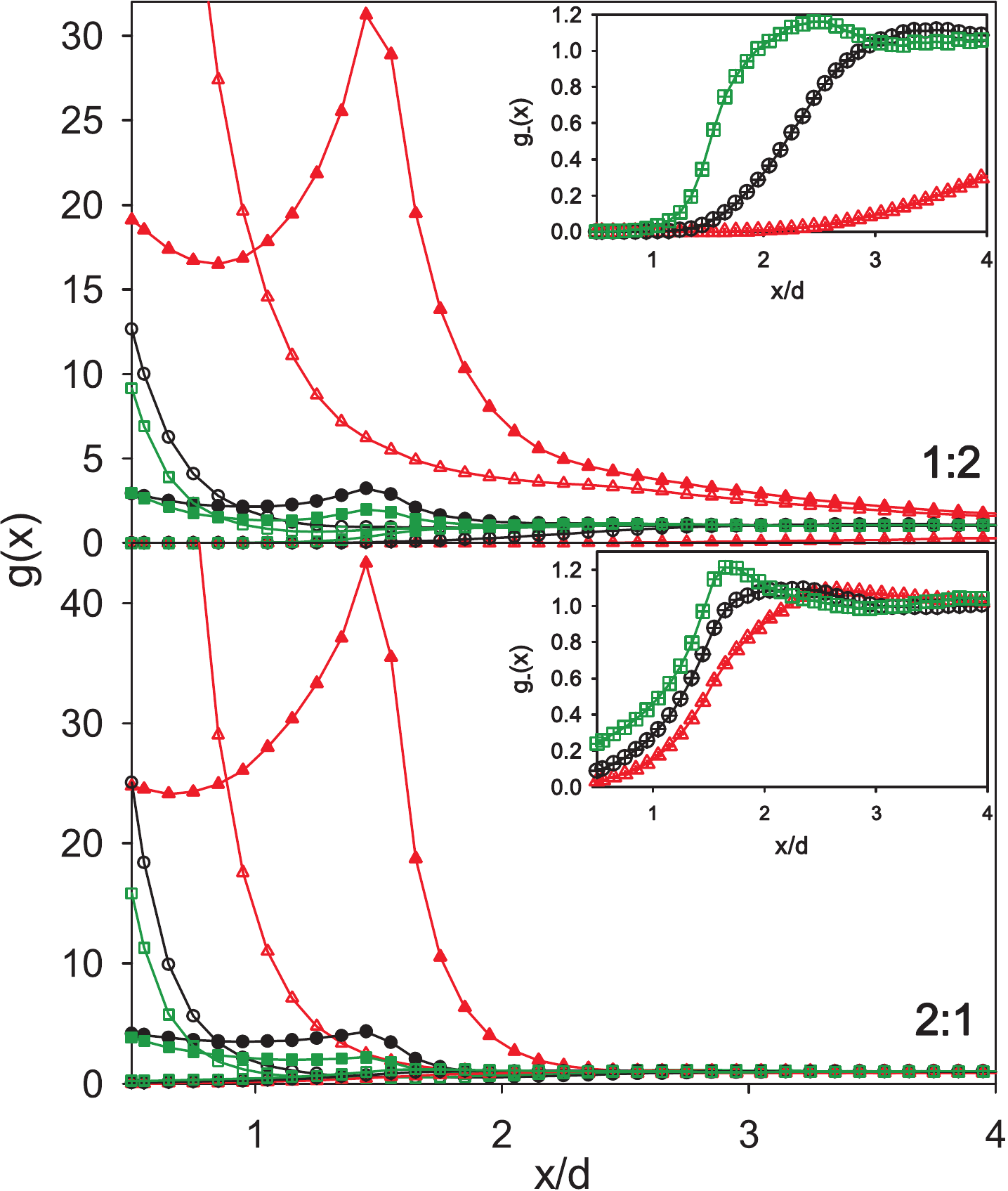}
}
\caption{(Color online) Dependence of the singlet distribution function, $g(x)$, of {charged dimers and spherical anions (see also the insets}) on the distance, $x/d$, from the electrode at the surface charge  $\sigma = -0.3$~C~m$^{-2}$ for electrolyte concentrations 0.10~M (triangles), 1.00~M (circles) and 2.00~M (squares) for ion valencies 1:2 (upper panel) and 2:1 (lower panel).
Empty symbols show the results for charged spheres, filled for the neutral spheres of dimers {and crossed for anions}.
\label{fig1}}
\end{figure}

Simulations were carried out for the asymmetric ion valencies 1:2 and 2:1 at three electrolyte concentrations 0.1, 1.0, and 2.0~M in the range of the electrode surface charges varying from $-1.0$ to $+1.0$~C~m$^{-2}$.
Diameters of positively charged and neutral spheres of dimers and of spherical anions are equal to $d =425$~pm.
The other physical parameters were temperature $T= 298.15$~K, and the relative permittivity of water, $\epsilon _\text{r}=78.5$.
Because of the anisotropic shape of cations, the simulations were expected to require significantly longer configurational sampling.
We sampled 2 billion configurations to obtain high precision averages.

Figure~\ref{fig1} shows the singlet distribution profiles of charged dimers {and spherical anions} for three electrolyte concentrations 0.1, 1.0 and 2.0~M at $\sigma = -0.30$~C~m$^{-2}$.
At this $\sigma$, a strong adsorption of dimers is observed, while anions are repelled from the vicinity of the electrode surface.
By contrast, at positive values of  $\sigma$, the adsorption of spherical anions occurs and now the dimer cations are repelled from the electrode. As a result, the internal structure of the distant
cation has little influence on the double layer structure with the dimer behaving like a spherical ion (see for example,
reference \cite{30.}).
The surface properties of spherical ions are well known, so we do not discuss them here.
The upper panel of figure~\ref{fig1} shows the results for the 1:2 electrolyte, while the lower one for the 2:1 case.
The density profiles of charged spheres are similar to those observed for spherical counter-ions.
A sharp peak corresponds to the contact distance, $d/2$.
As expected, its height decreases with an increasing electrolyte concentration.
The peak is higher and thinner for di-valent dimers as we have observed earlier \cite{30.}.
The influence of a neutral sphere is hardly visible.
The neutral sphere density profiles have two maxima.
The first corresponds to the contact distance while the second is at $x/d\approx  1.45$.
The height of the second maximum relative to the height of the first one increases with a decreasing electrolyte concentration.
The first maximum indicates that the large fraction of dimers take parallel orientation to the electrode surface.
Assuming the perpendicular orientation of a dimer with charged sphere adjacent the electrode surface, the centre of the neutral sphere is located at $x/d = 1.5$.
A good agreement of this prediction with the simulation results indicates that the perpendicular orientation is also very probable and its probability increases with a decreasing electrolyte concentration.
This conclusion confirms the mean orientation  $\langle \cos \theta \rangle$ results.

\begin{figure}[!t]
\centerline{
\includegraphics[width=0.5\textwidth]{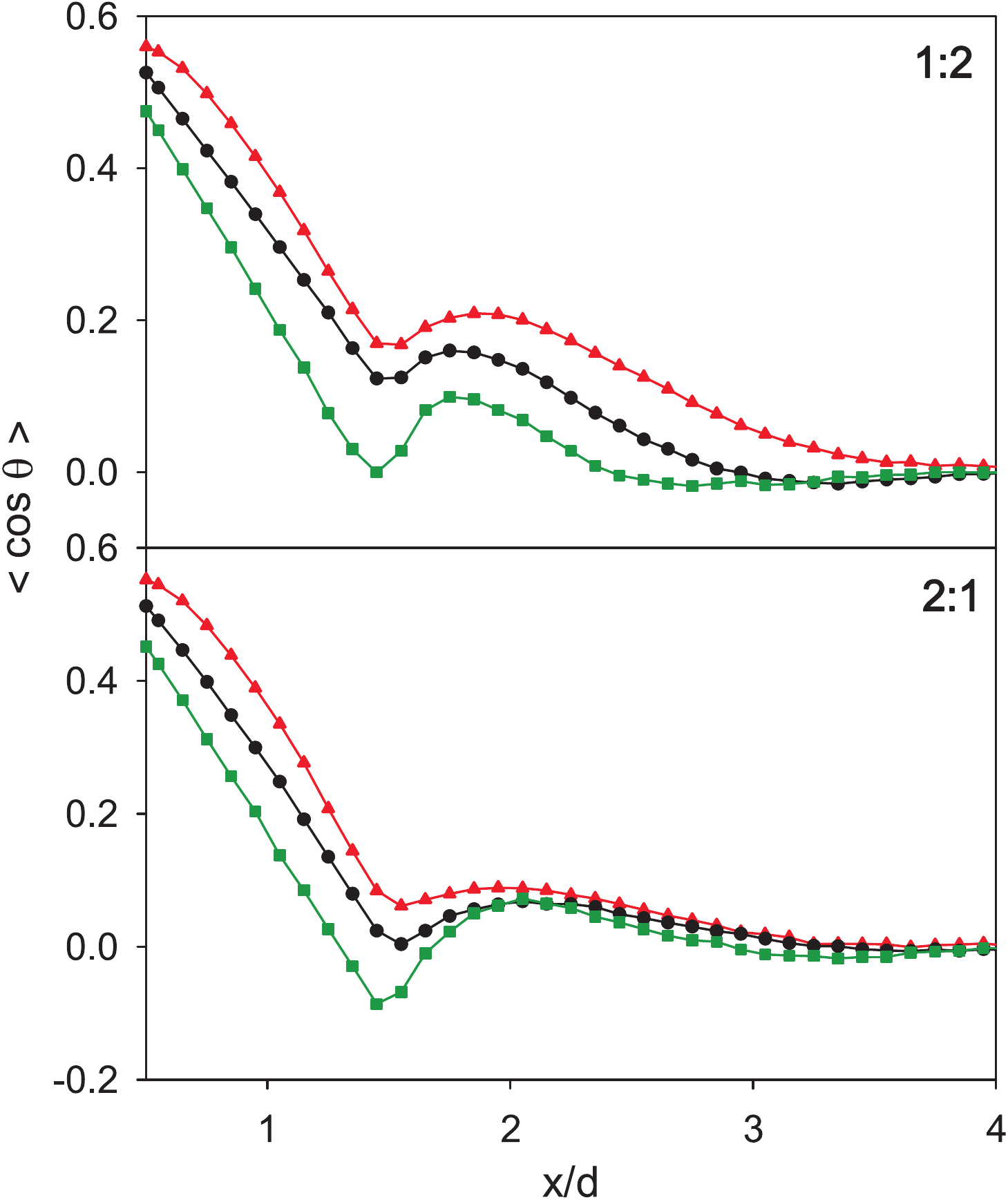}
}
\caption{(Color online) Dependence of the angular orientation function, $\langle \cos \theta \rangle$, of dimers on the distance, $x/d$. Parameters and symbols have the same meaning as in figure~\ref{fig1}.\label{fig2}}
\end{figure}

The mean orientation  $\langle \cos \theta \rangle$ profiles are shown in figure~\ref{fig2}.
The positive values of $\langle \cos \theta \rangle$ mean that dimers prefer the orientation with the charged sphere towards the electrode.
The function $\langle \cos \theta \rangle$ has a nearly linear course from the contact distance to the position of the second maximum on the neutral sphere density profile.
At this position, the $\langle \cos \theta \rangle$ function has a minimum.
The minimum is negative for higher electrolyte concentrations and larger dimer valencies.
At longer distances, the minimum transforms into a positive maximum.
After leaving the maximum, the $\langle \cos \theta \rangle$ function tends to zero and the orientation randomisation takes place.
Thus, the behaviour of $\langle \cos \theta \rangle$ suggests a generally perpendicular orientation of the dimers near the electrode with the charged head nearer to it than the neutral tail, a pattern that is consistent with our earlier studies.
The contact values of $\langle \cos \theta \rangle$  are nearly independent of the dimer valency, but at some distance from the electrode surface they are lower for the divalent dimers.
However, the course of curves remains similar.
In the vicinity of the electrode surface, the rotation of the neutral sphere around the charged one is hindered mainly by the hard wall of the electrode.
The wall effect is independent of electrolyte concentration.
The value of $\langle \cos \theta \rangle$ varies from 0.5 at a contact distance to 0 at $x/d = 1.5$.
The simulation contact results oscillate at around 0.5.
They are greater than 0.5 for ${c} = 0.1$~M and lower than 0.5 for ${c} = 2.0$~M.
It means that, as we have stated earlier, the low concentration electrolytes support the perpendicular orientation, while the electrolytes of high concentration enhance the parallel one.
In this range of distances, the rotation of the neutral sphere around the charged one is hindered by steric and electrostatic interactions with neighbouring ions, only.
The second maximum of $\langle \cos \theta \rangle$ shows that dimers form the second layer have their charged spheres oriented towards the electrode.
It is worth noting that this layer is not visible at the charged sphere density distributions for $c = 1.0$ and 2.0~M.

\begin{figure}[!t]
\centerline{
\includegraphics[width=0.55\textwidth]{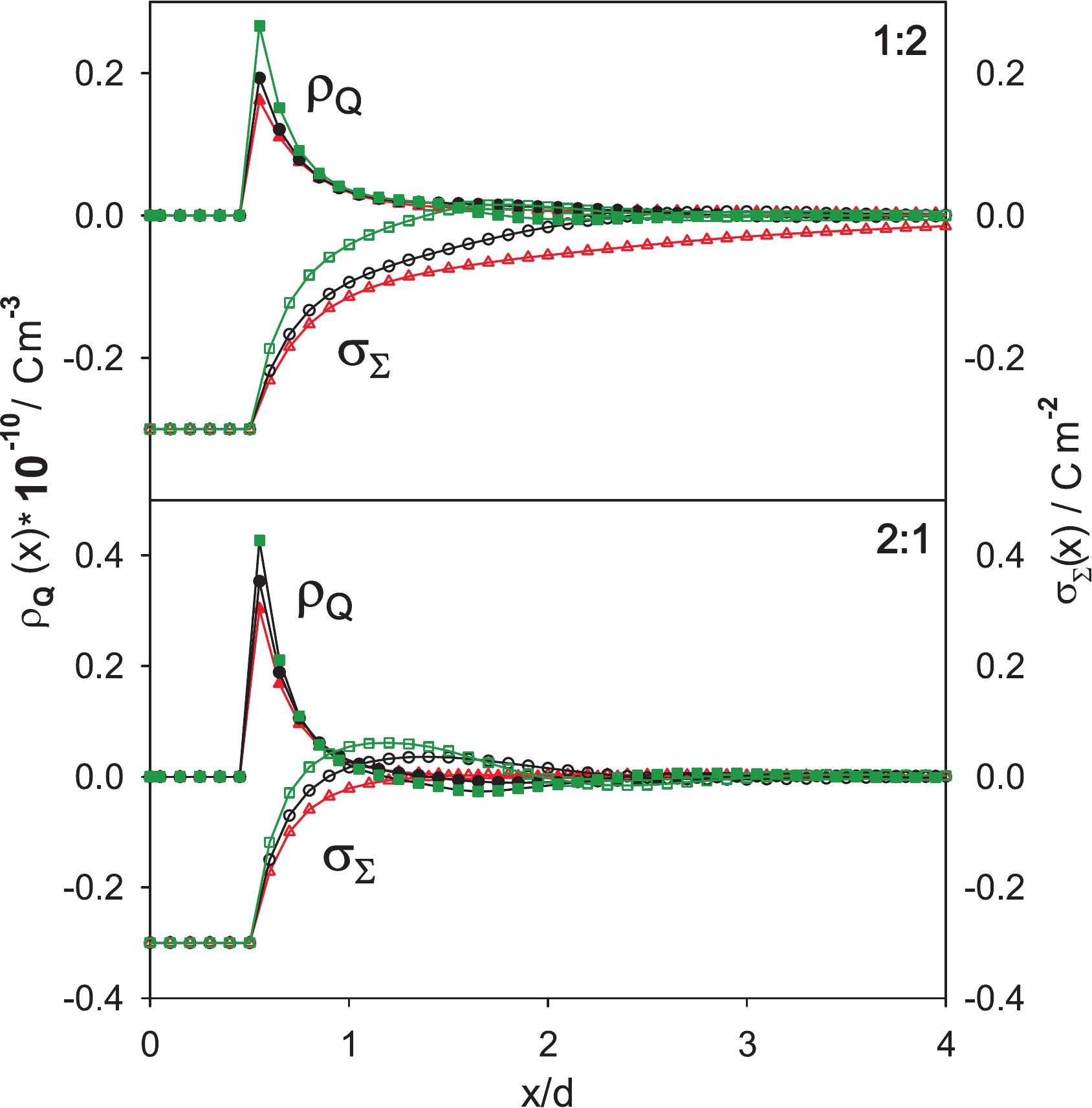}
}
\caption{{(Color online) Dependence of the local (volume) charge density $\rho_{Q} (x)$ (filled symbols) and the local net charge per unit area, $\sigma_{\Sigma} (x)$ (empty symbols) on the distance, $x/d$.  Parameters and symbols have the same meaning as in figure~\ref{fig1}.\label{fig3}}  }
\end{figure}

\begin{figure}[!b]
\centerline{
\includegraphics[width=0.48\textwidth]{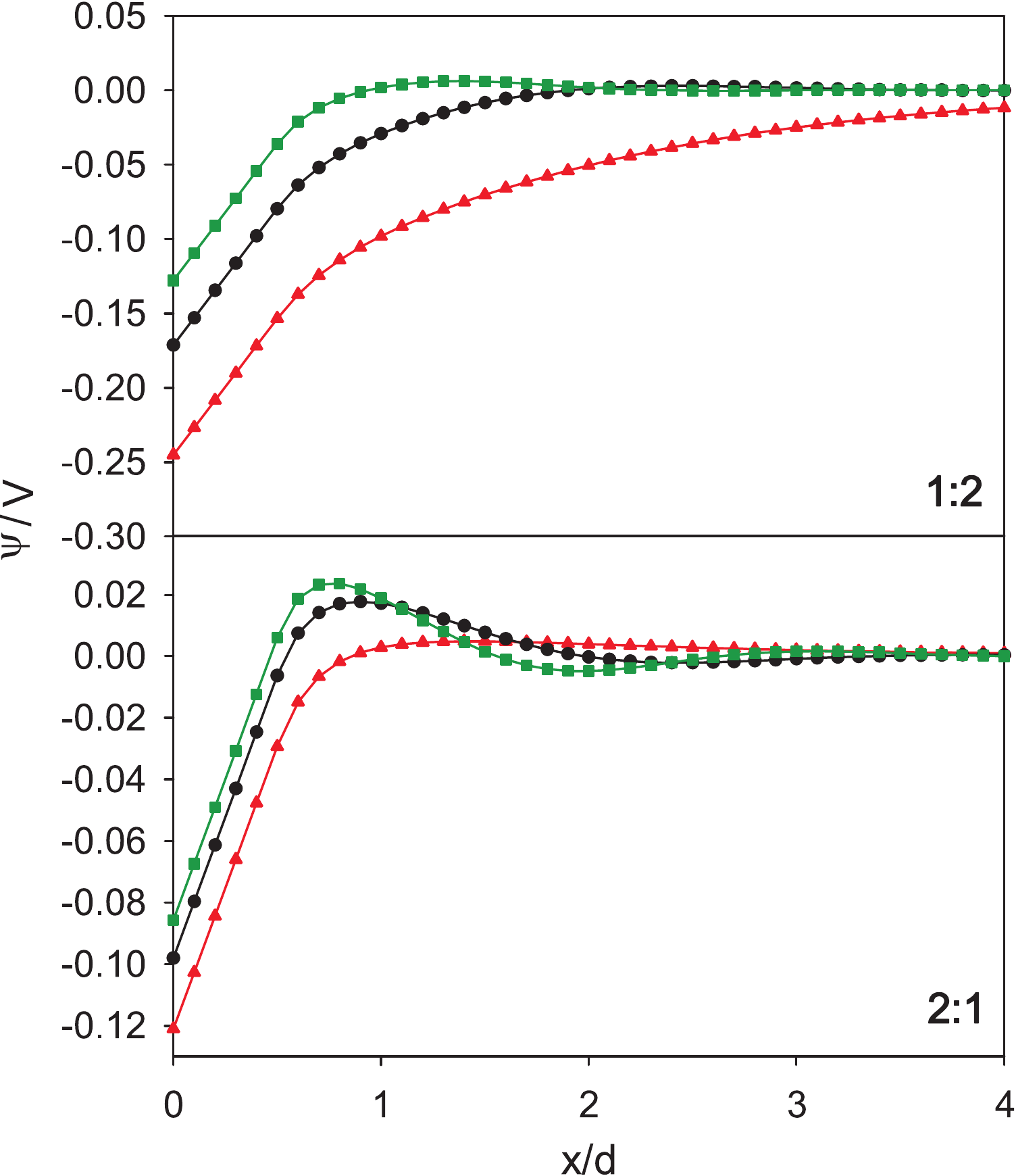} \hspace{10mm}
}
\caption{(Color online) Mean electrostatic potential $\psi$  as a function of  distance, $x/d$.  Parameters and symbols have the same meaning as in figure~\ref{fig1}. \label{fig4}}
\end{figure}

{Most of the anion distribution functions shown in figure~\ref{fig1} have a small maximum at $x/d > 1.5$ (see the insets).
This maximum suggests the onset of the CI and CR phenomena. As a case in point, we have
calculated the profiles of the local charge density, $\rho_{Q} (x)$, and the local net charge per unit area, $\sigma_{\Sigma} (x)$, which are presented in figure~\ref{fig3}.
It is seen that except the curve for the 1:2 electrolyte at concentration $c = 0.1$~M, which is monotonous, the remaining curves are all non-monotonous indicating the occurrence of the CI and CR phenomena.
The increase in electrolyte concentration or the presence of a higher valency counterion
intensifies these features. Indeed, at the same electrolyte concentration, the effect is
more pronounced for the 2:1 system than for the 1:2 system.
The CI effect is observed closer to the electrode surface than to the CR one.
The mechanism of CI and CR is not completely clear \cite{33.}. }

The mean electrostatic potential profiles calculated from equation (\ref{eq3})  are presented in figure~\ref{fig4}.
For the 1:2 electrolyte at ${c} = 0.1$ and 1.0~M, the potential is negative and the curves do not have any extrema.
The remaining curves have a positive maximum characteristic of divalent electrolytes \cite{38.}.
This maximum is related to the the CI effect.
With an increasing electrolyte concentration and dimer valency, the absolute value of the mean potential and the width of EDL decrease.

\begin{figure}[!t]
\centerline{
\includegraphics[width=0.49\textwidth]{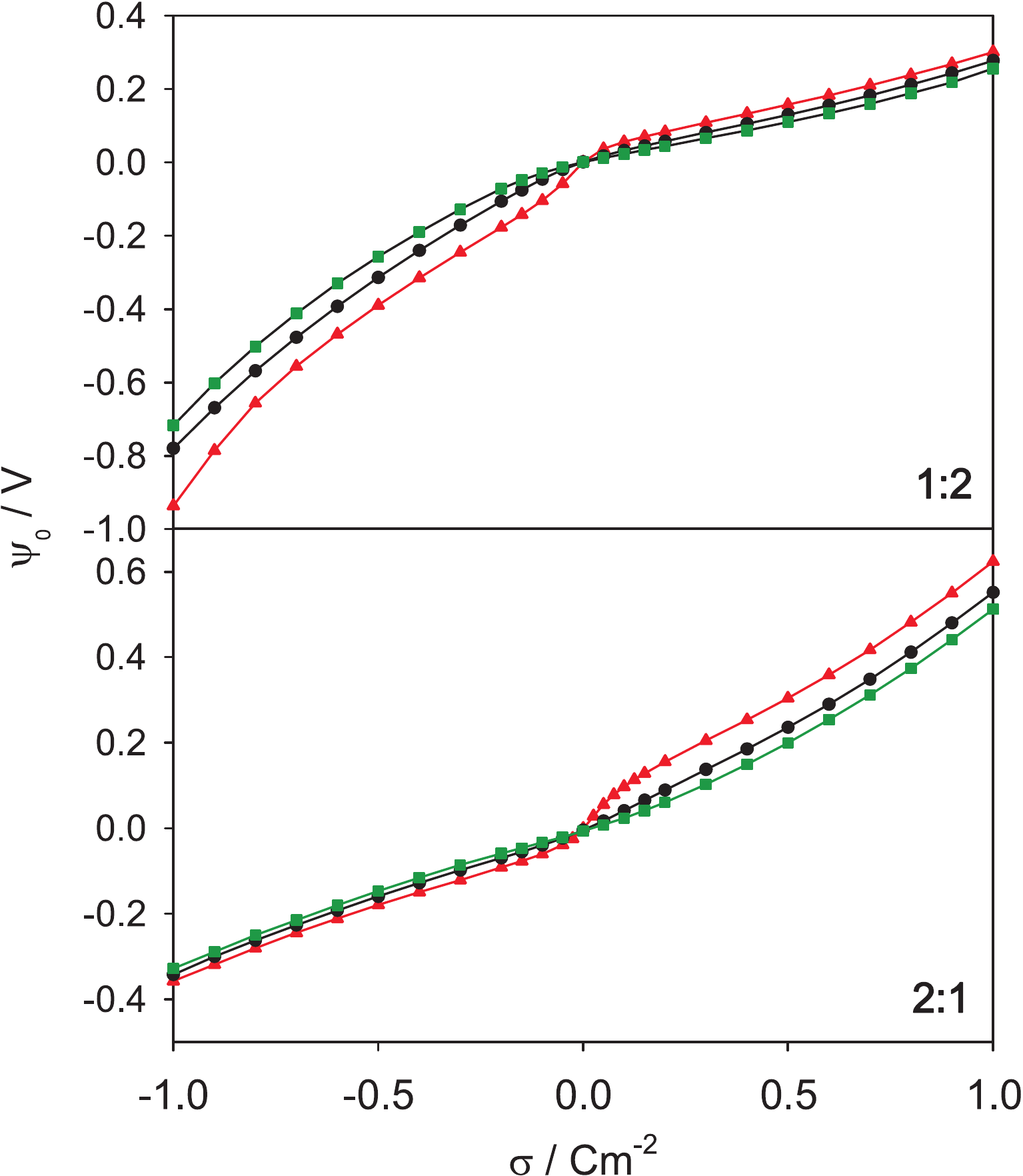}
}
\caption{(Color online) Mean electrostatic potential of the electrode, $\psi_{0}$  as a function of the surface charge density, $\sigma$. Symbols have the same meaning as in figure~\ref{fig1}. \label{fig5}}
\end{figure}

\begin{figure}[!b]
\centerline{
\includegraphics[width=0.49\textwidth]{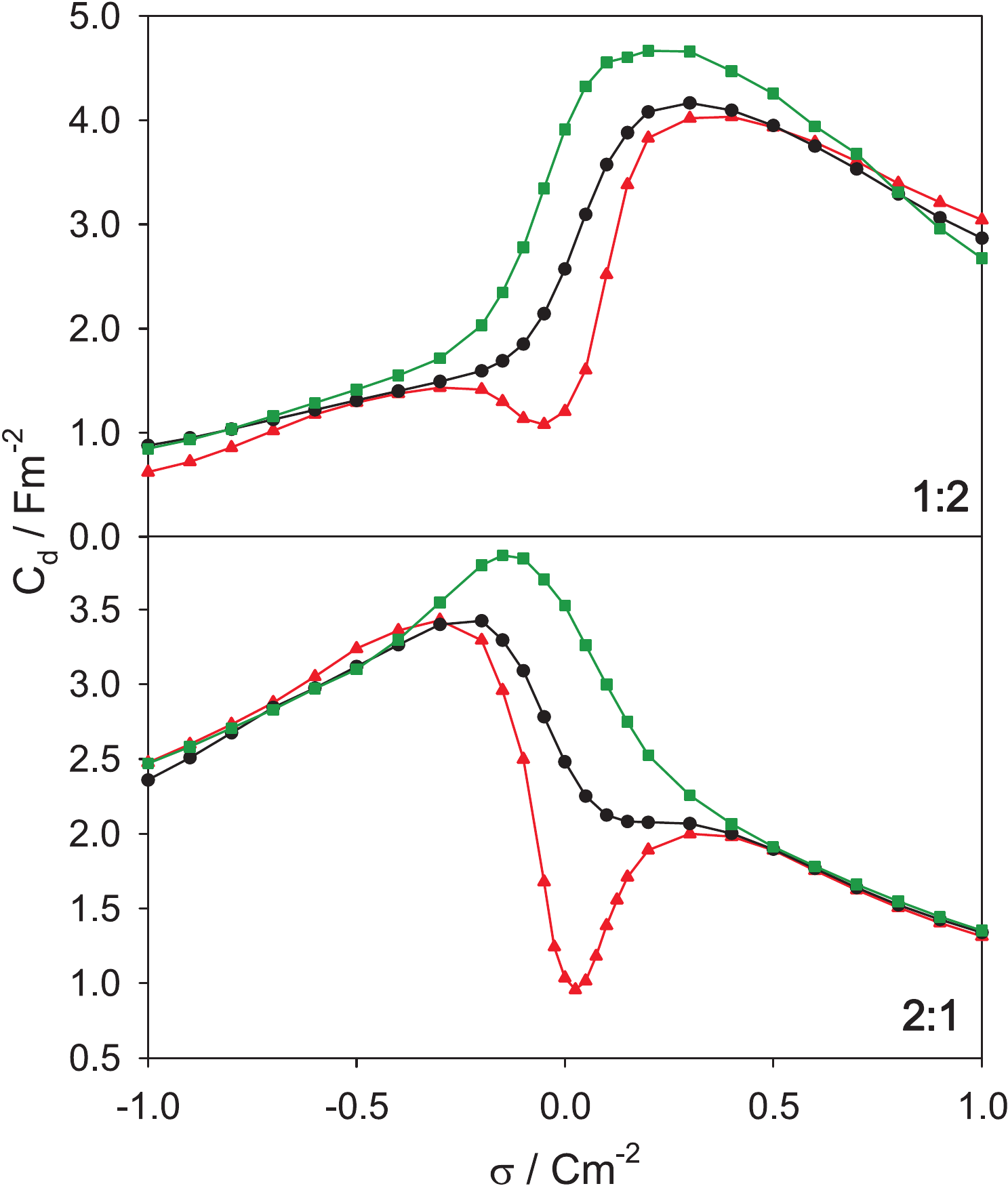}
}
\caption{(Color online) Differential capacitance, $C_d$, of the electrical double layer as a function of the surface charge density, $\sigma$. Symbols have the same meaning as in figure~\ref{fig1}. \label{fig6}}
\end{figure}

The dependence of the mean electrostatic potential of the electrode, $\psi(0)$, on the electrode charge density $\sigma$ calculated for the investigated electrolyte concentrations and on ion valencies is shown in figure~\ref{fig5}.
The curves intersect in the vicinity of $\sigma = 0$.
The slope of the curves in the range of adsorption of divalent ions is smaller than in the range of adsorption of monovalent ones.
The $\psi(0)$ results are used for the calculation of the differential capacitance of EDL.

Figure~\ref{fig6} shows the dependence of the differential capacitance, $C_\text{d}$, of EDL on the electrode charge density calculated for different asymmetric valencies and electrolyte concentrations.
For $c = 0.1$~M, the $C_\text{d}$ curves have a minimum corresponding to $\sigma\approx 0$ and surrounded by two asymmetric maxima.
An increase in the electrolyte concentration to $c= 2.0$~M transforms the minimum into a distorted maximum.
For the 1:2 electrolyte, the maximum is located at a positive $\sigma$, while for the 2:1 electrolyte, at a negative $\sigma$ value.
The Gouy-Chapman theory \cite{1.,2.} predicts that the differential capacitance curve of EDL has the parabola-like shape with a minimum at $\sigma = 0$.
Transition of the capacitance curve from that having a minimum to that having a  maximum with an increasing electrolyte concentration has been explained for the RPM electrolyte by an increase in the thickness of EDL due to the formation of bi- or multi-layer structures of counterions \cite{13.,39.}.
In the case of charged dimers, the increase in thickness of EDL is additionally caused by the neutral spheres of dimers which separate the two layers of charged spheres: adsorbed on the electrode surface and form the second layer.
Finally, the increase in valency of counterions makes the EDL thinner.
Explanation of the behaviour of $C_\text{d}-\sigma$ curve shown in figure~\ref{fig6} requires consideration of all these effects.

\section{Conclusions}

A charged dimer is a simple and useful model of molecules of charged surfactants and ionic liquids.
Charged surfactants are composed of a charged head and a neutral, {hydrophobic} tail.
In a charged dimer, the head is represented by a charged sphere while the tail by a neutral one.
Surfactants are diluted in a solvent (water).
{Therefore, more advanced investigation of surfactants modelled by a charged dimer should include solvent molecules.
The interaction of the charged head and the neutral tail with molecules of solvent depends strongly on the structural and polar properties of solvent molecules.
Presumably they can influence the density and orientation profiles. }
Ionic liquids are solvent free molten crystals composed of large organic ions with the electric charge located off the centre of the ion.
Charged dimers can model ionic liquids when their size is large and the concentration high.
Current investigation of EDL composed of ionic liquids as an electrolyte is extremely interesting.
However, the investigation of EDL formed by charged dimers by the method of  molecular simulations breaks down because of high density of a system.
The problem can be solved by replacing hard potentials with soft ones.

In this paper we have investigated the structural and thermodynamic properties of an electric double layer composed of a planar charged hard wall, charged dimers and spherical anions at different electrolyte concentrations and asymmetric ion valencies, using the grand canonical Monte Carlo simulation.
The density profiles of neutral spheres have two maxima corresponding to parallel and perpendicular orientations of the dimer against the electrode surface.
The height of the second maximum relative to the height of the first one increases with a decreasing electrolyte concentration.
Thus, we have concluded that the probability of the perpendicular orientation increases with a decreasing electrolyte concentration.

The variation in the electrolyte concentration and ion valency leads to a diversity of shapes of differential capacitance curves.
At low concentrations, the capacitance curves have a minimum surrounded by two asymmetric maxima.
With increasing concentrations, the minimum transforms into a distorted maximum.
For the 1:2 electrolyte, the maximum is located at positive $\sigma$, while for the 2:1 electrolyte, at a negative $\sigma$ value.

\section*{Acknowledgements}
Monika Kaja and Stanis{\l}aw Lamperski acknowledge the financial support from the Faculty of Chemistry,
Adam Micklewicz University of Pozna\'n.

\clearpage

\ukrainianpart

\title
{Вплив анізотропної форми іонів, асиметрії валентності та концентрації електроліту на структурні і термодинамічні властивості електричного подвійного шару}

\author{M. Кайя\refaddr{label1},
С. Ламперский\refaddr{label1},
В. Сильвестр-Алькантара\refaddr{label3},
Л.Б. Буян\refaddr{label3},
Д. Гендерсон\refaddr{label5}}

\addresses{
\addr{label1} Відділ фізичної хімії, Університет Адама Міцкевича в Познані,  Познань, Польща
%\addr{label2} Відділ фізичної хімії, Університет Адама Міцкевича в Познані,  Познань, Польща
\addr{label3} Лабораторія теоретичної фізики, фізичний факультет  Університету Пуерто Ріко, Пуерто Ріко
%\addr{label4} Лабораторія теоретичної фізики, фізичний факультет  Університету Пуерто Ріко, Пуерто Ріко
\addr{label5} Факультет хімії та біохімії, Університет Бригама Янга, Прово, США
}

\makeukrtitle

\begin{abstract}
Представлено результати моделювання Монте Карло симуляцій у великому канонічному ансамблі для електричного подвійного шару
поблизу плоскої зарядженої твердої поверхні, який складається з катіонів анізотропної форми та сферичних аніонів при різних концентраціях електроліту та для випадку асиметричної валентності. Катіони складаються з двох тангенційно зв'язаних твердих сфер однакового діаметру  ${d}$.
Одна сфера є зарядженою, а друга~--- нейтральною. Сферичні аніони~--- це заряджені тверді сфери діаметру ${d}$.
Розглядається асиметрія іонної валентності 1:2 і 2:1, причому іони знаходяться у розчиннику як діелектричне середовище при нормальній температурі.
Симуляції здійснювалися при концентрації електроліту: 0.1, 1.0 і 2.0~M.
Профілі розподілів електрод-іон, електрод-нейтральна сфера, середня концентрація димерів і середній електростатичний потенціал обчислені при фіксованому поверхневому заряді  $\sigma $, в той час як контактний потенціал електроду і диференційна ємність  представлені при заряді електроду, який змінюється. При збільшенні концентрації електроліта крива диференційної ємності змінюється від мінімуму, оточеного максимумами, до форми із спотвореним єдиним максимумом. Для електроліту 2:1, максимум розташований при невеликому від'ємному заряді, в той час як для електроліта 1:2~--- при невеликому позитивному значенні заряду.

\keywords заряджені димери, асиметрія валентності, електричний подвійний шар, Монте Карло моделювання у великому канонічному ансамблі
\end{abstract}

\end{document}